\documentclass[aps,prb,twocolumn,floats,showpacs]{revtex4}
\usepackage{epsfig}

\begin{document}
\title{Temperature-dependent quasiparticle band structure of the
  ferromagnetic semiconductor EuS}
\author{W.~M{\"u}ller}
\email[Email: ]{Wolf.Mueller@physik.hu-berlin.de}
\homepage{http://tfk.physik.hu-berlin.de} 
\author{W.~Nolting}
\affiliation{Lehrstuhl Festk{\"o}rpertheorie,
  Institut f{\"u}r Physik, Humboldt-Universit{\"a}t zu
  Berlin, Invalidenstra{\ss}e 110, 10115 Berlin, Germany}
\begin{abstract}
  We present calculations for the temperature-dependent electronic
  structure of the ferromagnetic semiconductor EuS. A combination of a
  many-body evaluation of a multiband Kondo-lattice model and a
  first-principles $T=0$--bandstructure calculation (\emph{tight-binding
    linear muffin-tin orbital (TB-LMTO)}) is used to get realistic 
  information about temperature- and correlation effects in the EuS energy
  spectrum. The combined method strictly avoids double-counting of any
  relevant interaction. Results for EuS are presented in terms of spectral
  densities, quasiparticle band structures and quasiparticle densities of
  states, and that over the entire temperature range.
\end{abstract}
\pacs{75.50.Pp, 75.10.-b, 71.20.-b, 71.15.Mb}
\maketitle

\section{Introduction}
Since the 1960s the europium chalcogenides EuX (X=O, S, Se, Te) have
attracted tremendous research activity, experimentally as well as
theoretically.\cite{Wachter79A,Zinn76,Nol79a} They are magnetic
semiconductors, which 
crystallize in the rocksalt structure with increasing lattice constants
(5 to 7{\AA}) when going from the oxide to the telluride. The Eu$^{2+}$
ions occupy lattice sites of an fcc structure so that each ion has
twelve nearest and six next-nearest Eu-neighbors.

As to their purely magnetic properties the EuX are considered almost
ideal realizations of the Heisenberg model for the so-called
\emph{local-moment magnetism}. Their magnetism is due to the
half-filled 4f shell of the Eu$^{2+}$. The 4f charge density
distribution is nearly completely located within the filled  $5s^{2}$
and $5p^{6}$ shells so that the overlap of 4f wavefunctions of adjacent
Eu$^{2+}$ ions is negligibly small. Hund´s rules of atomic physics can
be applied yielding an $^{8}S_{7/2}$ ground state configuration of the
4f shell. The $7\mu_{B}$ moments are exchange coupled resulting in
antiferromagnetic (EuTe, EuSe), ferrimagnetic (EuSe), and ferromagnetic
(EuO, EuS) orderings at low temperatures. The fact that the magnetic
contribution to the thermodynamics of the EuX is excellently described
by the Heisenberg Hamiltonian
\begin{equation}\label{eq:1}
  H=-\sum_{i,j}J_{ij}\mathbf{S}_{i}\cdot\mathbf{S}_{j}
\end{equation}
allows to test models of the microscopic coupling mechanism by direct
comparison to experimental data. There is convincing evidence that the
exchange integrals can be restricted to nearest ($J_{1}$) and next nearest
neighbors ($J_{2}$)\cite{BZDK80,BKZ84}. $J_{1}$ is positive (ferromagnetic)
decreasing from the oxide to the telluride. $J_{2}$ is negative, except
for EuO, where the antiferromagnetic coupling increases in magnitude
from the sulfide to the telluride. Liu and
coworkers\cite{Liu80,Liu83,LL83,LL84}  have
proposed an indirect exchange between the localized 4f moments mediated
by virtual excitations of chalcogenide-valence band (p) electrons into
the empty Eu$^{2+}$(5d) conduction bands together with a subsequent
interband exchange interaction of the d electron (p hole) with the
localized 4f electrons. Using this picture, very similar to the
Bloembergen-Rowland mechanism\cite{bloembergen95}, the calculated values
for $J_{1}$, 
$J_{2}$ agree nicely with experimental data, both in sign as well as in
magnitude, for EuO, EuS, and EuSe. The results are found by a
perturbative calculation of the indirect 4f-4f exchange interaction (\ref{eq:1})
with data from a \emph{linear combination of atomic orbitals (LCAO)} 
method as band structure input. The different distances of the 4f moments
do obviously create the different magnetic behavior of the EuX. For the
exchange integrals of the two ferromagnets one finds:\cite{BZDK80,BKZ84}
\begin{equation}\label{eq:2}
  \mathrm{EuO:}\quad \frac{J_{1}}{k_{\mathrm{B}}}=0.625\mathrm{K} ;\quad
  \frac{J_{2}}{k_{\mathrm{B}}}=0.125\mathrm{K} 
\end{equation}
\begin{equation}\label{eq:3}
  \mathrm{EuS:}\quad \frac{J_{1}}{k_{\mathrm{B}}}=0.221\mathrm{K} ;\quad
  \frac{J_{2}}{k_{\mathrm{B}}}=-0.100\mathrm{K} 
\end{equation}
Although in EuO the ferromagnetic interaction is more pronounced
($T_{\mathrm{C}}(\mathrm{EuO})=69.33$K;
$T_{\mathrm{C}}(\mathrm{EuS})=16.57$K)\cite{Wachter79A} a greater variety of 
experiments has been carried out with EuS than with EuO. The reason is
that single crystals as well as films with well defined
thicknesses\cite{goncharenko98,stachow-wojcik99:_ferrom_eus_pbs} can
better be prepared for EuS than for EuO. Apart 
from this, the ferromagnetism of EuS is interesting in itself for two
reasons. There are competing exchange integrals $J_{1}$ and $J_{2}$,
and the magnitude of the dipolar energy is comparable to the exchange
energy.

Besides the purely magnetic properties a striking temperature dependence
of the (empty) conduction bands has caused intensive investigation. This
was first detected for the ferromagnetic EuX as red-shift of the optical
absorption edge (4f-5d) upon cooling below
$T_{\mathrm{C}}$\cite{batlogg75}. The reason is 
an interband exchange coupling of the excited 5d electron to the
localized 4f electrons that induces the temperature dependence of the
localized moment system into the the empty conduction band states. A
further striking effect, which is due to the induced temperature
dependence of the conduction band states, is a metal-insulator
transition observed in Eu-rich EuO\cite{penny72,torrance72}. The Eu
richness manifests 
itself in twofold positively charged oxygen vacancies. One of the two
Eu$^{2+}$
excess electrons, which are no longer needed for the binding, is thought
to be tightly trapped by the vacancy. Because
of the Coulomb repulsion, the other electron occupies an impurity level
fairly close to the lower band edge. With decreasing temperature below
$T_{c}$ the band edge crosses the impurity level thereby freeing impurity
electrons. A conductivity jump as much as 14 orders of magnitude was
observed\cite{torrance72}. Other remarkable effects result from the
interaction of 
the band electron with collective excitations of the moment system. One
of these is the creation of a characteristic quasiparticle
(\emph{magnetic polaron}) which can be identified as a propagating
electron dressed by a virtual cloud of excited magnons.

In previous papers we have proposed a method for the determination of
the temperature dependent electronic structure of bulk
EuO\cite{schiller01:_temper_euo} as well as
EuO-films\cite{schiller01:_kondo}. The treatment is based on a
combination of a 
multiband Kondo-lattice model (MB-KLM) with first principles
tight-binding linear muffin-tin orbital (TB-LMTO) band structure
calculations. The many-body treatment of the (ferromagnetic) KLM was
combined with the first-principles part by strictly avoiding a
double-counting of any relevant interaction. The most striking result
concerned the prediction of a surface state for thin EuO(100) films, the
temperature shift of which may cause a surface halfmetal-insulator
transition\cite{schiller01:_predic_euo}. For low enough temperatures the
shift of the surface 
state leads to an overlap with the occupied localized 4f
states. Therefore, one can speculate that the resistivity of the
EuO(100) films might be highly magnetic field dependent, so that a
colossal magnetoresistance effect is to be expected.

In this paper we investigate in a similar manner the other ferromagnet
EuS, where we restrict ourselves first to the bulk material. We want to
derive the temperature dependent quasiparticle band structure (Q-BS), in
particular concentrating on those effects, which are due to a mutual
influence of localized magnetic 4f states and itinerant, weakly
correlated conduction band states. There was earlier work on the Q-BS of
bulk EuS\cite{borstel87,mathi00:_temper_eus}. In these papers, however,
an approach was employed 
that decomposes the Eu-5d band into five consecutive non-degenerate
subbands. For each of the subbands a single-band KLM was evaluated
therewith disregarding the full multiband-structure of the EuS
conduction band. Obviously this procedure leads to an overestimation of
certain correlation effects as a consequence of certainly too narrow
subbands.
We therefore use in this paper a multiband
4f-5d Kondo-lattice model to get reliably the temperature dependent Q-BS
of EuS with all correlation effects in a symmetry-conserving manner. Our
method combines a many-body analysis of the mentioned multiband-model
with a self-consistent LMTO band structure calculation.

Since the technical details can be taken from
\cite{schiller01:_temper_euo} we present in the 
following only the general procedure together with those aspects which
are vital 
for the understanding of the new EuS results. In Section 2  we formulate
the multiband Hamiltonian and fix its single-particle part by a
realistic band structure calculation. Furthermore, we describe the
parameter choice for the decisive interband exchange coupling. Section 3.1
is devoted to the local-moment ferromagnetism of EuS, while Section 3.2
repeats shortly how we solved the multiband Kondo-lattice model. In
Section 4 the electronic EuS structure is discussed in terms of
quasiparticle band structures and densities of states (Q-DOS) as well as
spectral densities, which are closely related to the angle and spin
resolved (inverse) photoemission experiment.

\section{Multiband Kondo-Lattice Model}

The complete model-Hamiltonian for a real substance with multiple
conduction bands consists of three parts:

\begin{equation}\label{eq:4}
  H=H_{5d}+H_{4f}+H_{df}  
\end{equation}

The first term contains the 5d conduction band structure of the
considered material as, e.g., EuS:

\begin{equation}\label{eq:5}
  H_{5d}=\sum_{i,j}\sum_{m,m'} T_{ij}^{mm'}c^{\dagger}_{im\sigma}c_{jm'\sigma}
\end{equation}

The indices m and m' refer to the respective 5d subbands, i and j to
lattice sites. c$_{im\sigma}^{\dagger}$ and c$_{im\sigma}$ are,
respectively, the creation and annihilation operator for an electron
with spin $\sigma$ ($\sigma=\uparrow, \downarrow$) from subband m at
lattice site $\mathbf{R}_{i}$. $T_{ij}^{mm'}$ are the hopping integrals,
which are to be obtained from an LDA calculation in order to incorporate
in a realistic manner the influences of all those interactions which are
not directly accounted for by our model Hamiltonian.

Each site $\mathbf{R}_{i}$ is occupied by a localized magnetic moment,
represented by a spin operator $\mathbf{S}_{i}$. It stems from the
half-filled 4f shell of the Eu$^{2+}$ ion, according to Hund's rule
being a pure spin moment of $7\mu_{B}$. The exchange coupled localized
moments are described by an extended Heisenberg Hamiltonian

\begin{equation}\label{eq:6}
  H_{4f}=-\sum_{ij}J_{ij}\mathbf{S}_{i}\cdot\mathbf{S}_{j} -D_{0}\sum_{i}\left(S_{i}^{z}\right)^{2}  
\end{equation}

In the case of EuS the exchange integrals $J_{ij}$ can be restricted to
nearest and next nearest neighbors (\ref{eq:3}). The non-negligible dipolar
energy in EuS is expressed by a single-ion anisotropy $D_{0}$.

The characterizing feature of the \emph{normal} single-band KLM, also
called \emph{s-f} or \emph{s-d model}, is an intraatomic exchange
between conduction electrons and localized spins. The form of the
respective multiband-Hamiltonian can be derived from the general on-site
Coulomb interaction between electrons of different subbands. It was
shown in \cite{schiller01:_temper_euo,schiller01:_kondo} that in the
special case of EuX (half-filled 4f shell, empty conduction band) the
interband exchange can be written as: 
\begin{eqnarray}\label{eq:7}
  \lefteqn{
  H_{df}=-\frac{1}{2}J\sum_{im}\Big(S_{i}^{z}
    \left(n_{im\uparrow}-n_{im\downarrow}\right)}
  \hspace{7em}\\
   &&
    +S_{i}^{+}c_{im\downarrow}^{\dagger}c_{im\uparrow}
    +S_{i}^{-}c_{im\uparrow}^{\dagger}c_{im\downarrow}\Big)
    \nonumber
\end{eqnarray}
J is the corresponding coupling constant, and furthermore:
\begin{equation}\label{eq:8}
  n_{im\sigma}=c^{\dagger}_{im\sigma}c_{im\sigma};\mbox{ }
  S^{\pm}=S^{x}\pm \mathrm{i}S^{y} 
\end{equation}
The first term in (\ref{eq:7}) represents an \emph{Ising type}
interaction 
while the two others refer to spin exchange processes. The latter are
responsible for some of the most typical KLM properties.

\begin{figure}[htbp]
  \centerline{\epsfig{file=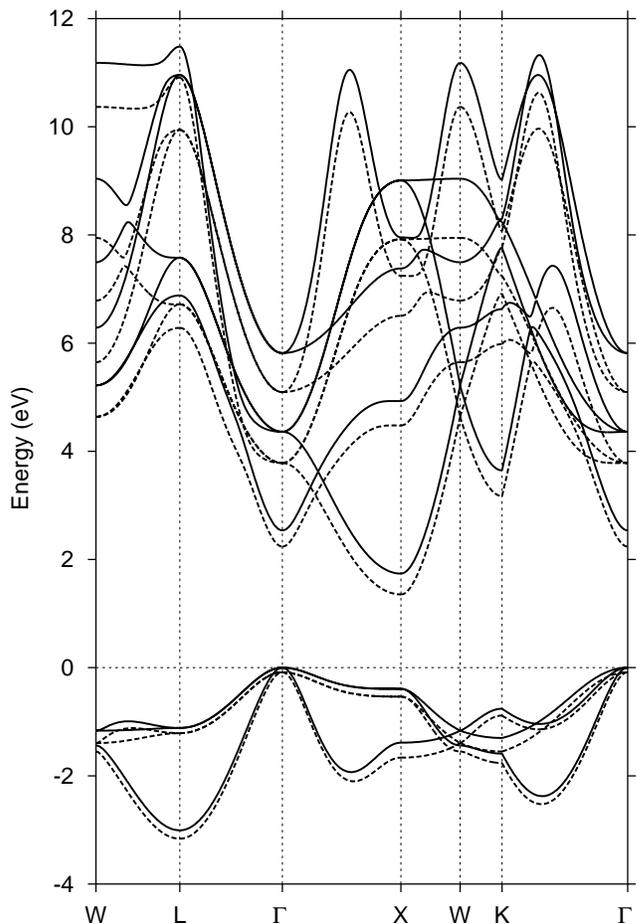,width=\linewidth}}
  \caption{Spin dependent (solid lines up-spin, broken lines down-spin)
    bandstructure of bulk EuS calculated within a TB-LMTO scheme with
    the 4f levels treated as core states. The energy zero coincides with
    the Fermi energy.}
  \label{fig:1}
\end{figure}

\begin{figure}[htbp]
  \centerline{\epsfig{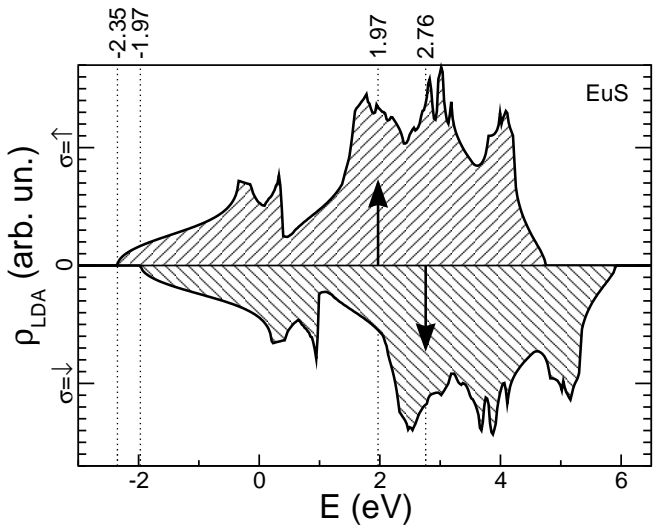}}  
  \caption{Spin-dependent density of 5d-states of EuS as function of
    energy (calculated within a TB-LMTO scheme). The numbers are (in eV)
    for the lower band edges and for the centers of gravity. The Fermi
    edge is below the 5d band (see Fig.~\ref{fig:1}).}
  \label{fig:2}
\end{figure}

In order to incorporate in a certain sense all those interactions, which
are not directly covered by the model Hamiltonian, we take the hopping
integrals from a band structure calculation according to the
TB-LMTO-atomic sphere program of
Andersen\cite{andersen75,andersen84}. 
In this method, the original Hamiltonian is transformed to a
tight--binding Hamiltonian containing nearest neighbor correlations,
only.
The transformation is obtained by linearly combining the original
muffin-tin orbitals to the short ranged tight--binding muffin-tin
orbitals. The evaluation is restricted to 5d bands, only.
LDA-typical 
difficulties arise with the strongly localized character of the 4f
levels. To circumvent the problem we considered the 4f electrons as core
electrons, since our main interest is focused on the reaction of the
conduction bands on the magnetic state of the localized moments. For our
purpose the 4f levels appear only as localized spins in the sense of
$H_{4f}$ in Eq.(\ref{eq:6}). 
Figure \ref{fig:1}  shows the calculated spin-dependent band
structure of EuS, where, of course, the 4f levels are missing. Clearly,
the conduction-band region is dominated by Eu-5d states. For our
subsequent model calculations it is therefore reasonable to restrict the
single-particle input from the band structure calculations to the Eu-5d
part, only. The low-energy part in Figure \ref{fig:1} belongs to the
S-3p states. 
For comparison we have also performed an LDA+U calculation which is able to
reproduce the right positions of the respective bands. However this
method suffers from the introduction of adjustable parameters $U$ and
$J$ being, therefore, no longer a ``first principles'' theory. The
results of the LDA+U calculation do not differ strongly from those of
the ``normal'' LDA, with the 4f electrons treated as core electrons. So
we have chosen the much simpler LDA calculation. Since we are mainly
interested in overall correlation and temperature effects, the extreme
details of the bandstructure are surely not so important.
In Figure \ref{fig:2}, from the same calculation, the LDA-density of states is
displayed. A distinct exchange splitting is visible which can be used to
fix the interband exchange coupling constant J in
Eq.(\ref{eq:7}). Assuming that 
an LDA treatment of ferromagnetism is quite compatible with the Stoner
(mean field) picture, as stated by several
authors\cite{janack76,poulsen76}, the $T=0$ 
splitting should amount to $\Delta E=JS$. Unfortunately, the
results in Figure \ref{fig:2} do not fully confirm this simple assumption but
rather point to an energy-dependent exchange splitting. The indicated
shifts of the lower edge and of the center of gravity lead to different
$J$ values:
\begin{equation}\label{eq:9}
  J=J(\mathrm{edge})=0.11\mathrm{eV};\quad
  J=J(\mathrm{center})=0.23\mathrm{eV}   
\end{equation}
In the following we will use both values for the a bit oversimplified
ansatz $H_{df}$ in Eq.(\ref{eq:7}) to compare the slightly different
consequences.

It is a well-known fact (see Figure \ref{fig:1} in
\cite{meyer01:_quant_kondo}  and further references
there) that the KLM can exactly be solved for the ferromagnetically
saturated ($T=0$) semiconductor. It is found that the $\uparrow$
spectrum is rigidly shifted towards lower energies by the amount of 
$\frac{1}{2}JS$, while the $\downarrow$ spectrum is remarkably deformed
by correlation effects due to spin exchange processes between extended
5d and localized 4f states. We cannot switch off the interband exchange
interaction $H_{df}$ in the LDA code, but we can exploit from the exact
$T=0$ result that it leads in the $\uparrow$ spectrum only to an
unimportant rigid shift. So we take from the LDA calculation, which
holds by definition for $T=0$, the $\uparrow$ part as the single-particle
input for $H_{5d}$ in Eq.(\ref{eq:5}). Therewith it is guaranteed that all the
other interactions, which do not explicitly enter the KLM operator
(\ref{eq:4}), 
are implicitly taken into account by the LDA-renormalized
single-particle Hamiltonian (\ref{eq:5}). On the other hand, a double
counting of 
any decisive interaction is definitely avoided.

\section{Model Evaluation}
\subsection{Magnetic Part}
\begin{figure}[htbp]
  \centerline{\epsfig{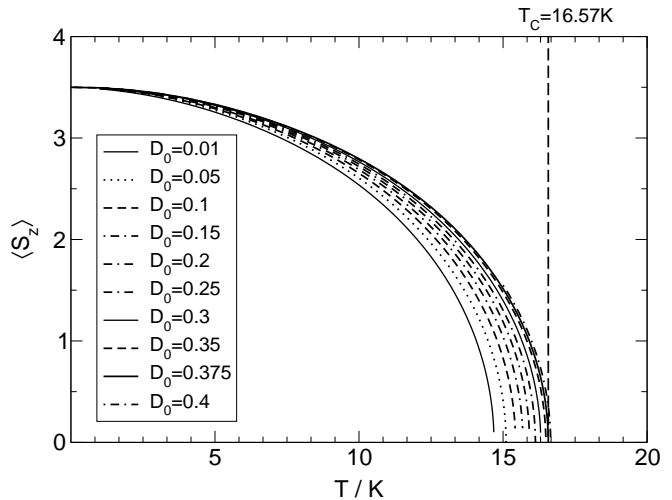}}
  \caption{4f-magnetization as a function of temperature for various
    values of the single--ion anisotropy $D_{0}$. The vertical broken
    line marks the used experimental value
    ($T_{\mathrm{C}}=16.57\mathrm{eV}$).  }
  \label{fig:3}
\end{figure}

Because of the empty conduction bands the magnetic ordering of the
localized 4f moments will not directly be influenced by the band
states. For the purely magnetic properties of EuS it is therefore
sufficient to study exclusively the extended Heisenberg-Hamiltonian
(\ref{eq:6}). While the exchange integrals $J_{ij}$ are known from spin wave
analysis (see Eq.(\ref{eq:3})), the single-ion anisotropy constant $D_{0}$ must
be considered an adjustable parameter. Via the magnon-Green function
\begin{equation}\label{eq:10}
  P_{ij}(E)=\langle\langle S_{i}^{+};S_{j}^{-}\rangle\rangle
\end{equation}
we can calculate all desired f spin correlation functions by evaluating
the respective equation of motion:
\begin{equation}\label{eq:11}
  EP_{ij}(E)=2\hbar^{2}\left<S_{i}^{z}\right>\delta_{ij}
  +\left\langle \left\langle
      \left[S_{i}^{+},H_{4f}\right]_{-};
      S_{j}^{-}\right\rangle\right\rangle_{E} 
\end{equation}
Evaluation of this equation of motion requires the decoupling of higher
Green functions, originating from the Heisenberg exchange term and the
anisotropy part in Eq.(\ref{eq:6}). For Green functions coming out of the
Heisenberg term we have used the so-called Tyablikow approximation,
which is known to yield reasonable results in all temperature
regions. For Green functions, which arise from the anisotropy term, we
use a decoupling technique proposed by Lines\cite{Lin67}. Details of the
method have been presented in a previous
paper\cite{schiller99:_thick_curie_heisen} on EuO. As result one 
gets the following well-known expression for the temperature dependent
local-moment magnetization:
\begin{equation}\label{eq:12}
  \left<S^{z}\right>=
  \hbar\frac{(1+\phi)^{2S+1}(S-\phi)+\phi^{2S+1}(S+1+\phi)}{\phi^{2S+1}-(1+\phi)^{2S+1}}
\end{equation}
$\phi$ can be interpreted as average magnon number:
\begin{equation}\label{eq:13}
  \phi=\frac{1}{N}\sum_{\mathbf{k}}\left(e^{\beta
      E(\mathbf{k})}-1\right)^{-1},  
\end{equation}
where $E(\mathbf{k})$ is the pole of the wave vector dependent Fourier
transform of $P_{ij}(E)$. Some typical magnetization curves are plotted
in Figure \ref{fig:3}. They differ by the value of the  anisotropy constant
$D_{0}$, which is still an undetermined parameter. When 
$D_{0}/k_{\mathrm{B}}$ increases from $0.01$K to $0.4$K
$T_{\mathrm{C}}$  rises from 
about $15$K to $16.9$K. Regarding that $J_{1}$, $J_{2}$ are derived from a
low-temperature spin-wave fit, the agreement between the calculated
$T_{c}$s and the experimental value of $16.57\mathrm{K}$\cite{Wachter79A} is
remarkably good  for almost all applied values of $D_{0}$, and best for
$0.375\mathrm{K}$. Since we
are interested above all in the electronic bulk band structure and its
temperature dependence, the actual numerical value of $D_{0}$ does not
play the decisive role. However, when treating systems of lower
dimensionality (films, surfaces), which is planned for a forthcoming
paper, then a finite $D_{0}$ will be the precondition for getting a
collective magnetic order of the spin
system\cite{MW66,gelfert01}. 

Together with (\ref{eq:12}) and (\ref{eq:13}) practically all local spin
correlations are derivable, as e.g.:
\begin{equation}\label{eq:14}
  \left<S^{-}S^{+}\right>=2\hbar\left<S^{z}\right> \phi
\end{equation}
\begin{equation}\label{eq:15}
  \left<(S^{z})^{2}\right> = \hbar^{2}S(S+1)\left<S^{z}\right>(1+2\phi)
\end{equation}
\begin{eqnarray}\label{eq:16}
  \left<(S^{z})^{3}\right>&=&\hbar^{3}S(S+1)\phi
  +\hbar^{2}\left<S^{z}\right>(S(S+1)+\phi)\nonumber\\
  &&-\hbar\left<(S^{z})^{2}\right>(1+2\phi)    
\end{eqnarray}
These and similar terms are responsible for the temperature dependence
of the electronic selfenergy.

\subsection{Electronic Part}
The inspection of the electronic part starts from the retarded Green
function
$\langle\langle c_{im\sigma};c_{jm'\sigma}^{\dagger}\rangle\rangle_{E}$
or its wave-vector dependent Fourier transform:
\begin{equation}\label{eq:17}
  \hat{G}_{\mathbf{k}\sigma}(E)=\frac{\hbar}{E\mathbf{1}
    -\hat{T}_{\mathbf{k}}-\hat{\Sigma}_{\mathbf{k}\sigma}(E)}  
\end{equation}
Here $\mathbf{1}$ represents the $(M\times M)$ identity matrix, where M is the
number of relevant subbands. The elements $T_{\mathbf{k}}^{mm'}$ of the
hopping matrix are the Fourier transforms of the hopping integrals in
Eq.(\ref{eq:5}), while the elements of the selfenergy matrix are introduced by
\begin{equation}\label{eq:18}
  \left\langle\left\langle\left[c_{im\sigma},H_{df}\right]_{-};
      c_{jm'\sigma}^{\dagger}\right\rangle\right\rangle_{E} \equiv
  \sum_{lm''}\Sigma_{il\sigma}^{mm''}(E)G_{lj\sigma}^{m''m'}(E)
\end{equation}
with subsequent Fourier transformation.

To get explicitly the selfenergy elements in Eq.(\ref{eq:18}) we evaluate the
commutator on the left-hand side what produces two higher Green
functions:
\begin{equation}\label{eq:19}
  \Gamma_{ikj\sigma}^{mm'}(E)= \left\langle\left\langle
      S_{i}^{z}c_{km\sigma};
      c_{jm'\sigma}^{\dagger}\right\rangle\right\rangle_{E}  
\end{equation}
\begin{equation}\label{eq:20}
  F_{ikj\sigma}^{mm'}(E)= \left\langle\left\langle
      S_{i}^{-\sigma}c_{km-\sigma};
      c_{jm'\sigma}^{\dagger}\right\rangle\right\rangle_{E}  
\end{equation}
$\Gamma$ arises from the \emph{Ising type} interaction in the d-f
interaction term (\ref{eq:7}) and F from the \emph{spin exchange}
partial operator 
$(S^{\uparrow,\downarrow}=S^{+,-})$:
\begin{equation}\label{eq:21}
  \left\langle\left\langle  \left[c_{im\sigma},H_{df}\right]_{-};
      c_{jm'\sigma}^{\dagger}\right\rangle\right\rangle_{E}=
  -\frac{1}{2}J\left(z_{\sigma}\Gamma_{iij\sigma}^{mm'}
    +F_{iij\sigma}^{mm'}\right) 
\end{equation}
$(z_{\sigma}=\delta_{\sigma\uparrow}-\delta_{\sigma\downarrow})$.
Exploiting already the fact that the EuS conduction band is empty we
encounter the following equations of motion of the higher Green
functions (\ref{eq:19}) and (\ref{eq:20}):
\begin{eqnarray}\label{eq:22}
  \lefteqn{\sum_{lm''}\left(E\delta_{kl}\delta_{mm''}
      -T_{kl}^{mm''}\right)\Gamma_{ilj\sigma}^{m''m'}(E)=} 
  \hspace{3em}\\ 
  &&\hbar\langle S^{z}\rangle\delta_{kj}\delta_{mm'}
    +\left\langle\left\langle S_{i}^{z}\left[c_{km\sigma},H_{df}\right]_{-};
      c_{jm'\sigma}^{\dagger}\right\rangle\right\rangle _{E}
  \nonumber
\end{eqnarray}
\begin{eqnarray}\label{eq:23}
  \lefteqn{\sum_{lm''}\left(E\delta_{kl}\delta_{mm''}-T_{kl}^{mm''}
    \right)F_{ilj\sigma}^{m''m'}(E)=} 
  \hspace{3em}\\
  &&\left\langle\left\langle(\delta_{\sigma\uparrow}
      S_{i}^{-}+\delta_{\sigma\downarrow}S_{i}^{+})
      \left[c_{km-\sigma},H_{df}\right]_{-};
      c_{jm'\sigma}^{\dagger}\right\rangle\right\rangle_{E}  
  \nonumber
\end{eqnarray}
On the right-hand side of these equations appear further higher Green
functions which prevent a direct solution and require an approximative
treatment. That shall be different for the non-diagonal terms ($i\neq k$)
and the diagonal terms ($i=k$), because the strong
intraatomic correlations due to the on-site interaction (\ref{eq:7}) have to be
handled with special care. For $i\neq k$ a self-consistent selfenergy
approach is applied, which has been tested in numerous previous
papers\cite{NRMJ97,NMJR96,schiller01:_temper_euo,schiller01:_kondo,meyer01:_quant_kondo,schiller99:_thick_curie_heisen}.
It simply consists in treating the 
commutators in (\ref{eq:22}) and (\ref{eq:23}), respectively, in formal
analogy to the definition equation (\ref{eq:18}) for the selfenergy: 
\begin{equation}\label{eq:24}
  \left\langle\left\langle S_{i}^{z}\left[c_{km\sigma},H_{df}\right]_{-};
      c_{jm'\sigma}^{\dagger}\right\rangle\right\rangle_{E} 
  \rightarrow\sum_{lm''}\Sigma_{il\sigma}^{mm''}\Gamma_{ilj\sigma}^{m''m'}(E)
\end{equation}
\begin{eqnarray}\label{eq:25}
  \lefteqn{
    \left\langle\left\langle(\delta_{\sigma\uparrow}
      S_{i}^{-}+\delta_{\sigma\downarrow}S_{i}^{+})
      \left[c_{km-\sigma},H_{df}\right]_{-}; 
      c_{jm'\sigma}^{\dagger}\right\rangle\right\rangle_{E}}
  \hspace{13em} 
  \\
  &&
  \rightarrow\sum_{lm''}\Sigma_{il-\sigma}^{mm''}F_{ilj\sigma}^{m''m'}(E)
  \nonumber
\end{eqnarray}
For the diagonal terms ($i=k$) a moment technique is used that takes the
local correlations better into account. For this purpose we explicitly
evaluate the commutators in Eqs.(\ref{eq:22}),(\ref{eq:23}) obtaining
then, as usual, 
further higher Green functions. In the first step these new functions
are reduced to a minimum number by exploiting that all functions, arising
from the \emph{Ising-equation} (\ref{eq:22}), can rigorously be
expressed by linear combinations of those which come out of the
\emph{spinflip-equation} (\ref{eq:23}). For a decoupling, the latter
are then written as linear combinations of simpler functions that are already
involved in the hierarchy of Green functions. The choice, which kind of
simpler functions enter the respective ansatz, is guided by exactly
solvable limiting cases (ferromagnetic saturation, zero-bandwidth limit,
$S=\frac{1}{2}$). We illustrate the procedure by a typical example:

$$D_{ij\sigma}^{mm'}(E)=\langle\langle\Delta_{i\sigma}c_{im\sigma};
c_{jm'\sigma}^{+}\rangle\rangle $$
\begin{equation}\label{eq:26}
  \Delta_{i\sigma}=\delta_{\sigma\uparrow}S_{i}^{-}S_{i}^{+}
  +\delta_{\sigma\downarrow}S_{i}^{+}S_{i}^{-} 
\end{equation}
For $S=\frac{1}{2}$ it holds rigorously:
\begin{equation}\label{eq:27}
  D_{ij\sigma}^{mm'}(E)=\frac{1}{2}G_{ij\sigma}^{mm'}(E)
  -(\delta_{\sigma\uparrow}
  -\delta_{\sigma\downarrow})\Gamma_{iij\sigma}^{mm'}(E) 
\end{equation}
This relation is valid for all temperatures. On the other hand, in
ferromagnetic saturation ($\langle S^{z}\rangle=S$) the same function reads for
all spin values:
\begin{equation}\label{eq:28}
  D_{ij\sigma}^{mm'}(E)=SG_{ij\sigma}^{mm'}(E) -(\delta_{\sigma\uparrow}
  -\delta_{\sigma\downarrow})\Gamma_{iij\sigma}^{mm'}(E) 
\end{equation}
Eqs. (\ref{eq:27}) and (\ref{eq:28}) clearly suggest the following
ansatz for the general case:
\begin{equation}\label{eq:29}
  D_{ij\sigma}^{mm'}(E)=\alpha_{\sigma}^{mm'}G_{ij\sigma}^{mm'}(E)
  +\beta_{\sigma}^{mm'}\Gamma_{iij\sigma}^{mm'}(E) 
\end{equation}
In order to fix the coefficients $\alpha_{\sigma}^{mm'}$ and
$\beta_{\sigma}^{mm'}$, we now calculate the first two spectral moments
of each of the three Green functions in (\ref{eq:29}), and that exactly and
independently from the respective Green function. The diagonal parts of
all other functions, appearing on the right-hand sides of (\ref{eq:22})
and (\ref{eq:23}), can be elaborated analogously.

By these manipulations we arrive at a closed system of equations for the
selfenergy matrix elements $\Sigma_{ij\sigma}^{mm'}(E)$, which can
numerically be solved. Via the spectral moments, used for the various
ansatze such as (\ref{eq:29}), a set of local spin correlations as those in
Eqs.(\ref{eq:12},\ref{eq:14},\ref{eq:15},\ref{eq:16}) enter the
procedure. They are mainly responsible for the temperature-dependence of
the electronic selfenergy.  

\begin{figure}[htbp]
  \centerline{\epsfig{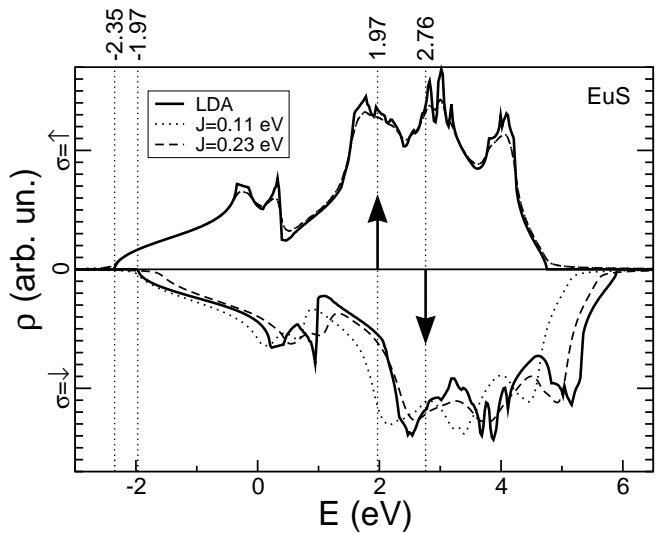}}
  \caption{The same as in Fig.~\ref{fig:2}, but in addition the $T=0$
    results of our combined many-body / first principles theory. The
    latter has been performed for two different values of the exchange
    coupling J (dotted line $J=0.11$eV; broken line $J=0.23$eV). The
    up-spin curves have been shifted both to coincide at the lower edge
    with the LSDA curve. That demonstrates that at $T=0$ correlation
    effects appear in the down--spectrum only.}
  \label{fig:4}
\end{figure}

To get a first impression of correlation effects in the electronic
structure of EuS we have evaluated our complex theory for T=0K. The
Q-DOS results are exhibited in Figure \ref{fig:4}. As explained and tentatively
justified before Eq.(\ref{eq:9}) we use two different values for the exchange
coupling J. The $\uparrow$ Q-DOS is unaffected by the actual value of J
and coincides with the respective LDA curve, when we compensate, as done
in Figure \ref{fig:4}, the unimportant rigid shift $(-\frac{1}{2}JS)$. So our
approach fulfills the exact $(T=0, \sigma=\uparrow)$-limit. The slight
deviations, seen in the upper part of Figure \ref{fig:4}, are exclusively due to
the numerical rounding procedure. \emph{A posteriori} this fact
demonstrates once more that our above-described method for
implementing the LDA input into the many-body model calculation
definitely circumvents the often discussed \emph{double counting
  problem}. As explained in Section 2 we succeeded in this respect
because for the special case of a ferromagnetically saturated
semiconductor the $\uparrow$ spectrum is free of correlation effects
which stem from the interband exchange $H_{df}$.

The lower half of Figure \ref{fig:4} demonstrates that correlation effects do
appear, even at $T=0$K, in the $\downarrow$ spectrum. Besides a band
narrowing, they provoke strong deformations and shifts with respect to
the LDA result. Here the influence of the different J values from
Eq.(\ref{eq:9}) is quite remarkable. For getting quantitative details of
the EuS-energy spectrum a proper choice of the exchange constant is obviously
necessary. The lower value $(J=0.11\mathrm{eV})$ is appropriate when we are
mainly interested in the lower band-edge region. However, in the middle of the
band, around the center of gravity, $J=0.23\mathrm{eV}$ is surely the better
choice.

\section{EuS Energy Spectrum}
\begin{figure}[htbp]
  \centerline{\epsfig{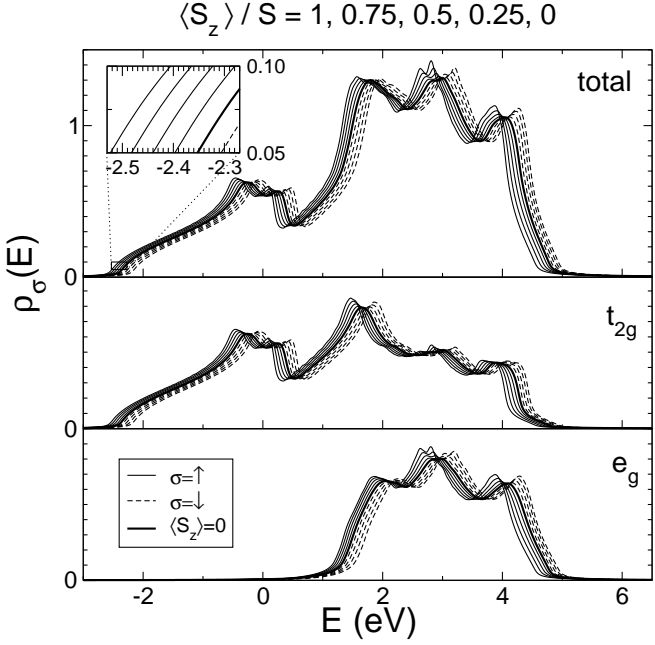}}
  \caption{Quasiparticle densities of states of the Eu--5d bands of bulk
    EuS as a function of energy for various temperatures. Solid lines
    for up--spin, broken lines for down--spin, thick line for
    $T=T_{\mathrm{C}}$ ($\langle S^z\rangle$=0). The outermost curves
    belong to $T=0$ ($\langle S^z\rangle/S=1$). They approach each other
    when increasing the temperature. The inset shows on an enlarged scale
    the temperature shift of the lower up--spin edge. Exchange coupling:
    $J=0.11$eV.}
  \label{fig:5}
\end{figure}

\begin{figure}[htbp]
  \centerline{\epsfig{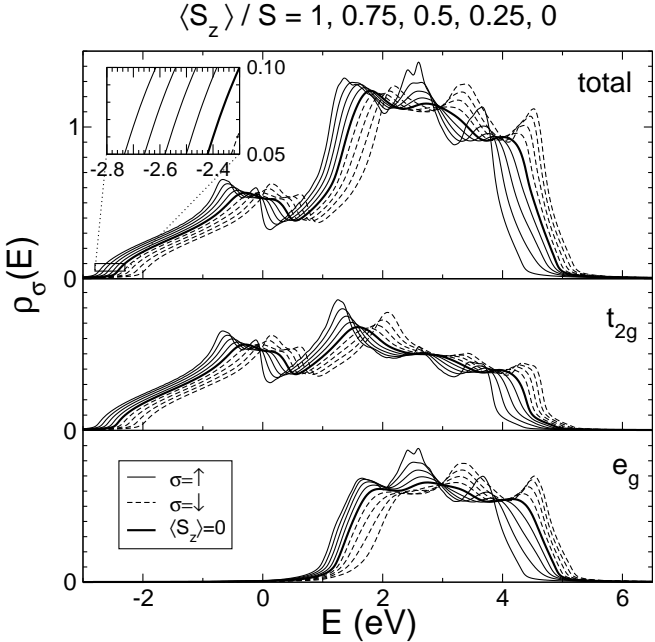}}
  \caption{The same as in Fig.~\ref{fig:5} but for $J=0.23$eV.}
  \label{fig:6}
\end{figure}

The main focus of our study is the temperature dependence of the
5d-energy spectrum of the ferromagnetic semiconductor EuS. The 5d bands
are empty, except for the single \emph{test electron}, so that the
T-dependence must be exclusively caused by the exchange coupling of the
band states to the localized \emph{magnetic} 4f states. Figure \ref{fig:5} and
Figure \ref{fig:6} display the quasiparticle densities of states for five different
4f magnetizations, i.e. five different temperatures (Figure \ref{fig:3}). The
Q-DOS in Figure \ref{fig:5} are calculated for $J=0.11\mathrm{eV}$. One sees that
the lower 
edge of the $\uparrow$ Q-DOS performs a shift to lower energies upon
cooling from $T=T_{\mathrm{C}}$ down to $T=0\mathrm{K}$. This explains
the famous red 
shift of the optical absorption edge for an electronic $4f-5d_{t_{2g}}$
transition, first observed by Busch and Wachter\cite{Wac64,BJW65}. We find a
shift of about $0.17\mathrm{eV}$ (see inset in Figure \ref{fig:5}),very
close to the 
experimental data\cite{Wachter79A,batlogg75}. This confirms once more that
$J=0.11\mathrm{eV}$ is a 
realistic choice for the exchange coupling constant as long as the lower
part of the 5d spectrum is under consideration.
Note, that our fitting procedure for the exchange constant $J$
(Fig.~\ref{fig:2}, Eq.~(\ref{eq:9})) does not predetermine the redshift.

Since we have taken into account for our calculation the full band
structure of the Eu-5d conduction bands, the symmetry of the different
Eu-5d orbitals is preserved. The 5d bands of bulk EuS are therefore
split into $t_{2g}$ and $e_{g}$ subbands (Figure \ref{fig:5}), where the $t_{2g}$
bands are substantially broader ($\sim 7\mathrm{eV}$) than the $e_{g}$ bands
($\sim 4\mathrm{eV}$). In a previous study of the temperature dependent
EuS-band structure\cite{borstel87} the Eu-5d complex was split into five
s-like bands by 
numbering for a given $\mathbf{k}$ vector the states from m=1 to m=5
according to increasing single-electron energies
$\epsilon_{m}(\mathbf{k})$. All $\mathbf{k}$ states with an energy
$\epsilon_{m}(\mathbf{k})$ then form the subband m. This simplified
procedure does not respect symmetries and neglects subband
hybridization, i.e. interband hopping $T_{ij}^{mm'}$ for $m\neq m'$. The
subband widths W turn out to be of order $1-3eV$ being therefore
distinctly narrower than those in Figure \ref{fig:5}. That has an important
consequence. Since correlation effects scale with the
\emph{effective(!)} exchange coupling $\frac{J}{W}$, they become for
the same J more pronounced in narrower bands. That is why we believe
that correlations are to a certain degree overestimated in
\cite{borstel87}.
As a consequence of the weaker effective coupling the appearance of
polaron-like quasiparticle branches is less likely in the present
investigation. 

In Figure \ref{fig:6} the Q-DOS is plotted for the stronger exchange coupling
$J=0.23\mathrm{eV}$, which should be more realistic for the middle part of the
spectrum, around the center of gravity. The temperature-influence on the
spectrum is more pronounced than for the weaker coupling in Figure
\ref{fig:5}. Strong deformations and shifts appear, being not at all rigid,
i.e. far beyond the mean field picture. However, not surprising, the
lower edge shift between $T=T_{\mathrm{C}}$ and $T=0\mathrm{K}$ comes
out too strong. The 
calculated red shift of $0.27\mathrm{eV}$ substantially exceeds the
experimental 
value of $0.167\mathrm{eV}$\cite{Wachter79A}. As mentioned above, the
lower part of the 
spectrum is better described with $J=0.11\mathrm{eV}$.

While the Q-DOS refers to the spin resolved, but angle averaged
(inverse) photoemission experiment, the $\mathbf{k}$ dependent spectral
density is the angle resolved counterpart. From
$S_{\mathbf{k}\sigma}(E)$ we derive the quasiparticle band structure
(Q-BS). 
\begin{figure}[htbp]
  \centerline{\epsfig{file=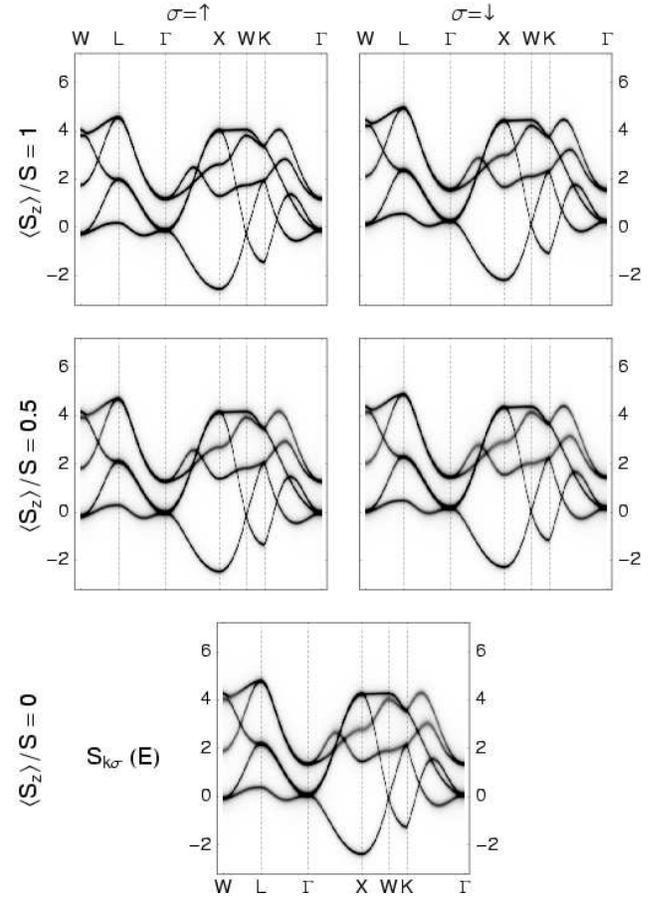,width=\linewidth,clip=}}
  \caption{Spin--dependent quasiparticle bandstructure of the Eu--5d
    bands of bulk EuS for different 4f magnetizations $\langle
    S^z\rangle/S$. Exchange coupling: $J=0.11$eV. }
  \label{fig:7}
\end{figure}
\begin{figure}[htbp]
  \centerline{\epsfig{file=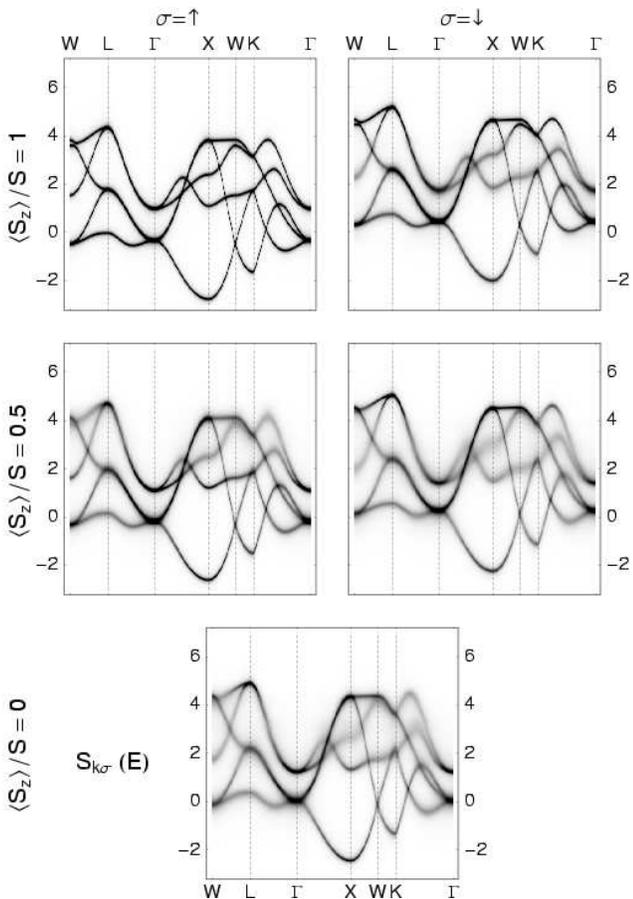,width=\linewidth,clip=}}  
  \caption{The same as in Fig.~\ref{fig:7}, but for $J=0.23$eV. }
  \label{fig:8}
\end{figure}
Figures \ref{fig:7} and \ref{fig:8} represent as density plots
the spectral density 
for some symmetry directions. The degree of blackening is a measure of
the magnitude of the spectral function. Figure \ref{fig:7} holds for
$J=0.11\mathrm{eV}$ 
and Figure \ref{fig:8} for $J=0.23\mathrm{eV}$. In both situations the $\uparrow$
spectrum 
in case of ferromagnetic saturation coincides with the dispersions
obtained from the LDA calculation. In the weak coupling case (Figure \ref{fig:7})
the temperature influence is mainly a shift of the total spectrum. The
induced exchange splitting reduces with increasing T and disappears at
$T=T_{\mathrm{C}}$. Correlation effects are more clearly visible in the case of
$J=0.23\mathrm{eV}$ (Figure \ref{fig:8}). They manifest themselves above
all in lifetime 
effects. Great parts of the dispersions are washed out because of magnon
absorption (emission) of the itinerant $\uparrow (\downarrow)$ electron
with simultaneous spinflip. In ferromagnetic saturation a $\uparrow$
electron has no chance to absorb a magnon because there does not exist
any magnon. Therefore the dispersion appears sharp representing
quasiparticles with infinite lifetime. On the other hand, the
$\downarrow$ electron has even at $T=0\mathrm{K}$ the possibility to emit a
magnon becoming then a $\uparrow$ electron. Therefore correlation
effects are already at $T=0\mathrm{K}$ present in the $\downarrow$
spectrum. For finite temperatures, finite demagnetizations, magnons are
available and absorption processes provoke quasiparticle damping in the
$\uparrow$ 
spectrum, too. The overall exchange splitting reduces with increasing
temperatures, until in the limit $T\rightarrow T_{c}$
($\langle S^{z}\rangle\rightarrow 0$) the vanishing 4f magnetization
removes the induced spin asymmetry in the 5d subbands.
\begin{figure}[htbp]
  \centerline{\epsfig{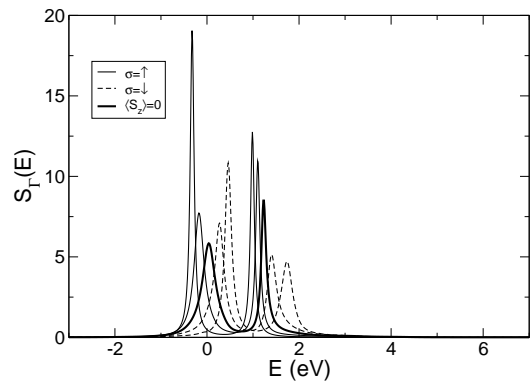}}  
  \caption{Spin--dependent spectral density $S_{\mathbf{k}\sigma}$ of
    the Eu--5d states of bulk EuS as function of energy for the same 4f
    magnetizations as in Fig.~\ref{fig:8}. Solid lines:
    up--spin; broken lines: down--spin; thick lines: $T=T_{\mathrm{C}}$
    ($\langle S^z\rangle=0)$). Exchange coupling: $J=0.23$eV;
  $\mathbf{k}=\Gamma$.}
  \label{fig:9}
\end{figure}
\begin{figure}[htbp]
  \centerline{\epsfig{file=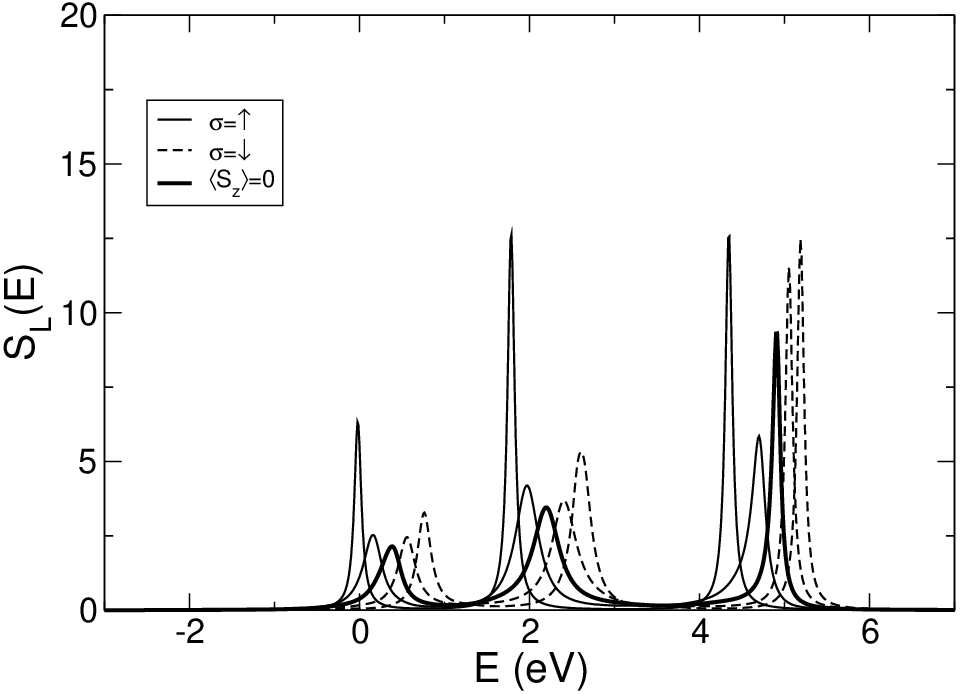,width=0.8\linewidth,clip=}}  
  \caption{The same as in Fig.~\ref{fig:9} but $\mathbf{k}=\mathrm{L}$.}
  \label{fig:10}
\end{figure}
\begin{figure}[htbp]
  \centerline{\epsfig{file=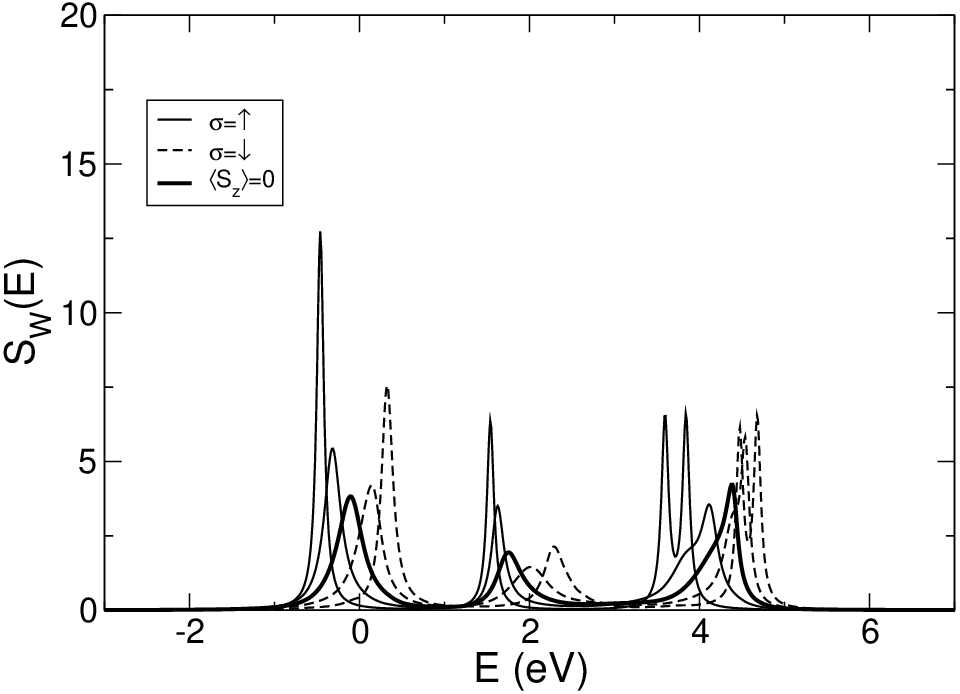,width=0.8\linewidth,clip=}}  
  \caption{The same as in Fig.~\ref{fig:9} but $\mathbf{k}=\mathrm{W}$.}  
  \label{fig:11}
\end{figure}
\begin{figure}[htbp]
  \centerline{\epsfig{file=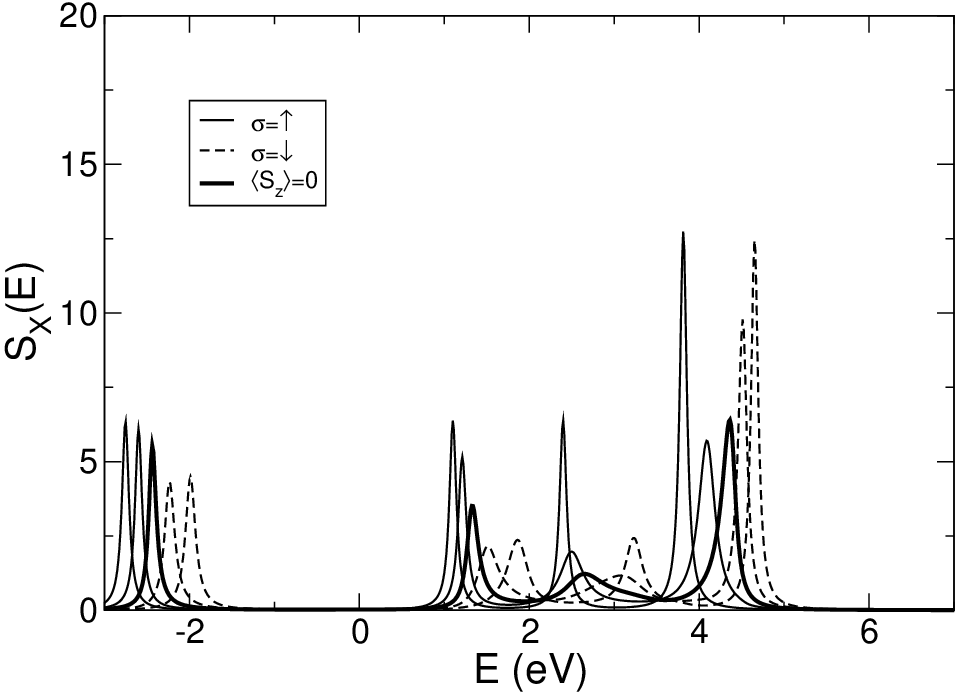,width=0.8\linewidth,clip=}}  
  \caption{The same as in Fig.~\ref{fig:9} but $\mathbf{k}=\mathrm{X}$.}  
  \label{fig:12}
\end{figure}

For a better insight into the temperature-behavior we have plotted in
Figs.~\ref{fig:9}--\ref{fig:12} for four $\mathbf{k}$--points from the
Brillouin zone ($\Gamma$, L, W, X) the energy dependence of the spectral
density, and that for the same three temperatures as in
Fig.~\ref{fig:8}. For the $\Gamma$--point  we expect two structures
according to the twofold (e$_{\mathrm{g}}$) and threefold
(t$_{\mathrm{2g}}$) degenerate dispersions. As can be seen in
Fig.~\ref{fig:9} well defined quasi particle peaks appear with
additional spin split below $T_{\mathrm{C}}$. The exchange splitting,
introduced via the interband coupling to the magnetically active 4f
system, collapses for $T\rightarrow T_{\mathrm{C}}$ (\emph{Stoner--like
behavior}). Obviously, quasiparticle damping increases with increasing
temperature. Similar statements hold for the spectral density at the
L--point. In accordance with the quasiparticle bandstructure in
Fig.~\ref{fig:8} three structures appear, the upper two being twofold
degenerate (Fig.~\ref{fig:10}). Interesting features can be observed at
the W--point (Fig.~\ref{fig:11}). At $T=0$ four sharp peaks show up in
the $\uparrow$--spectrum, and, though already strongly damped, the same
peak--sequence comes out in the $\downarrow$--spectrum.
The exchange splitting amounts to about $0.8$--$0.9$eV. With increasing
temperature damping leads to a strong overlap of the two upper peaks,
which are no longer distinguishable. 

At the X--point the spin resolved
spectral density exhibits four clearly separated structures, where the
upper belongs to a twofold degenerate dispersion (Fig.~\ref{fig:12}). In the
two middle structures, at least the $\downarrow$--peaks are so strongly
damped that they certainly will not be observable in an inverse
photoemission experiment. Altogether, the 5d--spectral densities of the
ferromagnetic semiconductor EuS exhibit drastic
temperature--dependencies, what concerns the positions and the widths of
the quasiparticle peaks.

\section{Conclusions}
\label{sec:Conc}
We presented a method of calculating the temperature dependent
bandstructure for the ferromagnetic semiconductor EuS. The essential
point is the combination of a many body evaluation of a proper theoretical
model with a ``first principles'' band structure calculation. The model
of choice is the ferromagnetic Kondo--lattice model, which describes the
exchange interaction between localized magnetic moments and itinerant
conduction electrons. For a realistic treatment of EuS we have extended
the conventional KLM to a multiband version to account for orbital
symmetry. Intraband-- and interband--hopping integrals have been
extracted from a \emph{tight--binding linear muffin--thin orbital} band 
structure calculation to incorporate besides the \emph{normal}
single--particle energies the influence of all those interactions which
are not directly covered by the KLM--Hamiltonian. By exploiting an exactly
solvable limiting case of the KLM this combination of first--principles and
model--calculations could be done under strict avoidance of a
double--counting of any relevant interaction.

The many--body part of the procedure was performed within a
moment--conserving interpolation method that reproduces exactly
important rigorous limiting cases of the model. The resulting electronic
selfenergy carries a distinct temperature--dependence, which is mainly due
to local 4f spin correlations. Since the band is empty, the KLM reduces
to a simple Heisenberg model as long as the purely magnetic
4f properties, as e.g.\ the mentioned 4f spin correlations, are
concerned. The result is a closed system of equations which can be
solved numerically for all quantities we are interested in.

We have demonstrated the temperature--dependence of the energy spectrum
of the ferromagnetic semiconductor EuS in terms of the 5d spectral
density and 5d  quasiparticle density of states. Peak positions and peak
widths determine energies and life--times of quasiparticles,
which have been gathered in special quasiparticle band structures. 
A striking temperature--dependence of the \emph{empty}
5d--bands is observed which is induced by the magnetic 4f--state. A
well--known experimental confirmation of the $T$--dependence is the
``red shift'' of the optical absorption edge, uniquely related to the
shift of the lower 5d--edge for decreasing temperature from
$T=T_{\mathrm{C}}$ down to $T=0$K. The induced exchange splitting at
$T=0$ collapses for $T\rightarrow T_{\mathrm{C}}$ \emph{Stoner--like},
but with distinct changes in the quasiparticle damping (lifetime). All these
temperature--dependent band effects should be observable by use of
inverse photoemission.

We expect further insight into interesting physics by a forthcoming
extension of our method to finite band occupations (Gd, Gd--films). The
respective single--band model version has already been presented in
previous papers \cite{NRMJ97,rex99:_temper_gadol}. In particular a
\emph{modified RKKY} has been used\cite{santos02} for a selfconsistent
inclusion of the magnetic properties of the not directly coupled 
local moments. A further implementation of disorder in the local moment
system will allow to treat diluted magnetic semiconductors such as
Ga$_{1-x}$Mn$_x$As and therefore contribute to the hot topic
\emph{spintronics}.

\subsection*{Acknowledgment}
Financial support by the ``Sonderforschungsbereich 290'' is gratefully
acknowledged. 

\bibliographystyle{revtex}

\end{document}